# Quasiparticle Interference Evidence of the Topological Fermi Arc States in Chiral Fermionic Semimetal CoSi

Qian-Qian Yuan,[1,2]* Liqin Zhou,[3,4]* Zhi-Cheng Rao,[3,4] Shangjie Tian,[7] Wei-Min Zhao,[1,2] Cheng-Long Xue,[1,2] Yixuan Liu,[7] Tiantian Zhang,[3,4] Cen-Yao Tang,[3,4] Zhi-Qiang Shi,[1,2] Zhen-Yu Jia,[1,2] Hongming Weng,[3,5,6]† Hong Ding,[3,5] Yu-Jie Sun,[3,5,6]† Hechang Lei,[7] Shao-Chun Li[1,2]†

[1] *National Laboratory of Solid State Microstructures, School of Physics, Nanjing University, Nanjing 210093, China*

[2] *Collaborative Innovation Center of Advanced Microstructures, Nanjing University, Nanjing 210093, China*

[3] *Beijing National Laboratory for Condensed Matter Physics and Institute of Physics, Chinese Academy of Sciences, Beijing 100190, China*

[4] *University of Chinese Academy of Sciences, Beijing 100049, China*

[5] *CAS Center for Excellence in Topological Quantum Computation, University of Chinese Academy of Sciences, Beijing 100190, China*

[6] *Songshan Lake Materials Laboratory, Dongguan, Guangdong 523808, China*

[7] *Department of Physics and Beijing Key Laboratory of Opto-electronic Functional Materials & Micro-nano Devices, Renmin University of China, Beijing 100872, China*

\* These authors contributed equally to this work.

† Corresponding authors:

scli@nju.edu.cn; yjsun@iphy.ac.cn; hmweng@iphy.ac.cn

**Chiral Fermions existed as quasiparticles in solid state feature the surface "Fermi arc" states, which connect the surface projections of the bulk chiral nodes with opposite Chern numbers. The surface Fermi arc is experimentally accessible as one of the most significant signature to manifest the nontrivial bulk topology. Aside from the Weyl nodes as firstly uncovered with Chern number C = ±1, chiral fermions carrying larger Chern number in CoSi family candidates have been theoretically proposed. Distinctly, the bulk chiral nodes in CoSi are enforced at high symmetric momenta in Brillouin zone by nonsymmorphic crystalline symmetry, and thus an extensive Fermi arc traversing the whole Brillouin zone is expected. Herein, we use scanning tunneling microscopy / spectroscopy (STM / STS) to investigate the quasiparticle interference (QPI) at various terminations of CoSi single crystal. The observed surface states exhibit the chiral fermion-originated characteristics. For instance, they are found to reside on (001) and (011) but not (111) surfaces with π-rotation symmetry, to spiral with energy, and to disperse in a wide energy range from ~ -200 mV to ~ +400 mV. Owing to the high energy and space resolution, a spin-orbit coupling induced splitting of up to ~ 80 mV is identified for the first time. Our experimental observations are corroborated by density functional theory (DFT) simulation, and thus provide a strong evidence that CoSi hosts the unconventional chiral fermions and extensive surface Fermi arc states.**

**INTRODUCTION**

Recently, great progress has been achieved in condensed matter physics in search of the analog of the elementary particles as described in high-energy physics. The three types of fundamental fermions—Dirac, Weyl and Majorana—have been discovered in solids, in the form of low-energy fermionic excitations near the topologically or symmetrically protected band crossing(*1-17*). Because these Fermionic excitations are constrained by the crystalline symmetry much lower than the Poincare symmetry in high-energy physics, new types of Fermions that have no high-energy counterparts have also been proposed and found in condensed matter materials(*18-30*), including spin-3/2 Rarita-Schwinger Weyl (RSW) excitations(*26, 27*), three-fold nexus fermions(*22, 24*), spin-1 Weyl fermions(*28*), double Weyl fermions(*29*) and double Dirac fermions(*30*) etc. These unconventional chiral fermions may exhibit fantastic physical properties, such as the helical surface states(*31, 32*), unusual magnetotransport(*33-35*), and the circular photogalvanic effect(*36, 37*), etc.

The chiral crystalline family of transition metal silicides, including CoSi, RhSi, RhGe, and CoGe, has been recently proposed as ideal candidates to host unconventional chiral Fermion quasiparticles through ab-initio calculations(*38-40*). They are expected to have numbers of advantages against the previously explored Weyl semimetals. For example, multiple types of topological chiral nodes coexist and locate close to the Fermi energy; there is no trivial bands crossing the Fermi energy and the transport properties are expectedly dominated by the chiral fermions(*38, 39*); as protected by the nonsymmorphic crystalline symmetry and time-reversal symmetry, the Fermi arc is rather extensive, in contrast to the short ones reported in previous Weyl semimetals(*7-12*). These extraordinary chiral fermions in CoSi family are the spin-1 Weyl fermions of three-fold degeneracy and the double Weyl fermions of four-fold degeneracy when spin-orbit coupling (SOC) is ignored in spinless case. If SOC is considered, the above spin-1 Weyl fermion evolves into a spin-3/2 RSW fermion of four-fold degeneracy (with spin degree of freedom) and the double Weyl fermion becomes a double spin-1 Weyl fermions of six-fold degeneracy (again with spin degree of freedom)(*38, 39*). In the absence of symmetry constraint, these chiral nodes in CoSi family are not necessary to be with equal energy, and in a wide energy window can one in principle observe the Fermi arc states. Bulk band structure and surface states of CoSi have been characterized by angle-resolved photoemission spectroscopy (ARPES)(*41-44*). However, a full understanding of the exotic surface Fermi arc is still lacking, even though which plays an indispensable role in the determination of the bulk topology and the nontrivial properties. In this work, we systematically investigated the surfaces of CoSi single crystal by using high resolution STM/STS technique, and proved the surface Fermi arc states through quasiparticle interference (QPI) measurement which is corroborated by DFT simulation.

**RESULTS AND DISCUSSIONS**

The crystal structure of CoSi belongs to the non-symmorphic space group $P2_13$ (No. 198) with the lattice constant of a = b = c = 4.45 Å (*45*), as illustrated in Fig. 1A. The unit cell of CoSi contains four Co and four Si atoms, with each Co (Si) atom covalently bonded with six nearest neighboring Si (Co) atoms. In the Co (Si) - terminated (001) surface, the Co (Si) atom and the underlying Si (Co) atom located out of center form a zig-zag atomic chain. The reciprocal Brillouin zone of CoSi is sketched in Fig. 1B, where the time reversal invariant momenta are marked. The projected surfaces crossing the Γ point for (001), (011) and (111) are highlighted in different colors. DFT calculation indicates that the bulk energy band of CoSi contains high-fold degenerate crossings at Γ and R, as shown in Fig. 1C. The band crossing at Γ point forms the spin-3/2 RSW Fermion node of four-fold degeneracy, and that at R point forms the double spin-1 Weyl node of six-fold degeneracy. The two types of chiral fermions are illustrated in Fig. 1D. Due to the no-go theorem, the RSW fermion node carries the chiral charge of +4 and the double spin-1 Weyl fermion -4. Therefore, if Γ / R are projected to different momenta at the surface BZ, for instance, $\bar{\Gamma}$ / $\bar{M}$ on the (001) and $\bar{\Gamma}$ / $\bar{X}$ on the (011) surfaces as illustrated in Fig. 1E and F, extensive

Fermi arcs are formed as the chiral conducting states between them. The inclusion of SOC leads to the doubling of the Fermi arcs. However, for the case of (111) surface, the Γ and R points are projected to the same momentum at the center of the surface BZ, and their opposite Chern numbers of ±4 thus overlap with each other. As a result, the topologically protected Fermi arc from the surface states connecting the two opposite chiral nodes is not observable, as illustrated in Fig.1G.

Atomically flat surfaces of various terminations of CoSi single crystal can be achieved after cycles of Argon ion sputtering and annealing in ultrahigh vacuum. Figure 2A shows for example the STM topographic image of the (001) surface. The step height, as shown in the inset to Fig. 2A, is measured to be ~ 4.5 Å, consistent with the lattice constant along [001] direction. The step height of ~ 2.25 Å is also occasionally observable which corresponds to the distance between two adjacent atomic planes, i.e., half size of the unit cell, see Supplementary Information fig. S1. The high resolution STM image in Fig. 2B clearly displays the square lattice of a = 4.45 Å. The zig-zag chain structure formed by Co-Si bonding, as shown in the inset to Fig. 2B, further confirms that the exposed surface is (001) oriented. The other two surfaces of (011) and (111) are also geometrically confirmed, as shown in Fig. 2C and D and Supplementary Information fig. S1. The atomically-resolved images displayed in Fig. 2C show the rectangle lattice of a = 4.45 Å and b = 6.27 Å, for the (011) surface and in Fig. 2D the hexagonal lattice of a = 6.27 Å for the (111) surface.

Differential conductance dI/dV spectrum (reflecting the local density of state, LDOS) taken on the (001) surface of CoSi shows the non-vanishing states in the whole bias range of ±0.2 V, indicating the (semi-) metallic nature, as shown in Fig. 2E. Intriguingly, the LDOS exhibits a singular peak near the Fermi energy. It is generally believed that either the flat dispersion of a bulk band or the surface states can result in an enhancement in LDOS. In order to figure out the origin of this singular peak, we took the dI/dV measurement on the (011) and (111) surfaces as well for comparison, as shown together in Fig. 2E and Supplementary Information fig. S2.

One can see that the dI/dV spectra taken on the (011) and (001) surfaces are obviously distinct from that on the (111) surface. The former two display similar LDOS peaks near Fermi energy, but in the latter, the intensity of the peak is completely suppressed. According to previous DFT calculation and ARPES measurement(*38, 39, 41-44*), the flat bulk band is projected to all of the three surfaces. However, as illustrated in Fig .1, there exist only surface Fermi arc states on the (001) and (011) surfaces, and no such surface states on the (111) surface. Therefore we deduced that the LDOS peak observed on the (001) and (011) surfaces is most likely related with the presence of the topological surface Fermi arc states. To further manifest it, the LDOS including /excluding the surface states are calculated for the projected surfaces, as displayed in Fig. 2F. There is indeed no prominent peak feature near Fermi energy in the bulk states. But the prominent feature due to the existence of the surface states on the (001) surface qualitatively agrees with the measured dI/dV spectra. It is noteworthy that the dI/dV spectrum taken on (111) surface is in good agreement with the calculated LDOS without surface states, as shown in Fig. 2F.

Quasiparticle interference (QPI) measurements were carried out at the three surfaces of (001), (011) and (111) to further characterize the surface Fermi arc states. The QPI patterns, as derived from the fast Fourier transform (FFT) of dI/dV maps measured in real space, can reveal the wave vectors of the electron's elastic scattering, and thus the dispersive information of the electron bands. It has been widely employed in manifesting the existence of topological surface states in topological insulators and semimetals(*46-50*). The FFT-STS maps taken on the (001) surface are shown in Fig. 3A and B. More information can be found in fig. S3. Extensive features can be recognized near Fermi energy. As marked in Fig. 3A, four prominent eye-shaped features (yellow arrows) are located at the corners of the first Brillouin zone and the two crescent moon-shaped features (black arrow) reside near the Bragg points. These features are beyond the scattering between bulk electron (hole) pockets as usually expected as compact pockets in QPI, as shown in Fig. 3E and F. It is obvious that these extensive features do not appear equivalently around the four orthogonal Bragg peaks (or along the orthogonal high-symmetry directions) and without mirror symmetry, but exist with π-rotation symmetry, suggesting that the patterns are likely from the chiral surface states.

To quantitatively understand the origin of the scattering channels, we performed DFT-generated QPI simulation. Both QPI simulations for (001) surface including and excluding surface states are presented in Fig. 3C and D, E and F, respectively. Surprisingly, similar eye-shaped patterns in the simulation including surface states are also identified, which are mainly resulted from the scattering channel of $q_1$ between surface Fermi arc states, as marked in Fig. 3G. The scattering channel $q_2$ between surface states leads to a similar pattern as crescent moon. Additionally, the calculated feature is also of π-rotation symmetry, in agreement with the experimental observation. It is noteworthy that the Fermi arc contour depends sensitively on the real atomic geometry of the surface. The QPI pattern may look different as well if spin texture of the Fermi arc is considered. In any case, the qualitative consistency between experiment and simulation undoubtedly reveals the existence of surface Fermi arc states. Instead, bulk-state associated QPI features, i.e., the expected pocket-like patterns as illustrated in Fig. 3E and F, are not well distinguishable, which may be due to either the weakly scattering of the 3D-like bulk states, or being overwhelmed by the surface state scattering. The presence of surface Fermi arc states is also supported by the QPI data taken on the (011) and (111) surfaces. Figure 3I and L show the QPI results taken on the (011) and (111) surfaces, and Fig. 3J-K and M-N the corresponding QPI simulations and calculated constant energy contours. More dI/dV maps taken on the (011) and (111) surfaces can be found in Supplementary Information figs. S4 to S7 and Movie M1. As marked in Fig. 3I for (011) surface, the arc-shaped feature can be clearly observed connecting two Bragg points, and shows π-rotation symmetry. This arc-shaped feature can be reproduced in the corresponding simulated QPI (Fig. 3J), which is associated with the scattering between the topological surface states as marked by the black dashed lines in Fig.3K. However, no similar extensive feature is observed on (111) surface, as shown in Fig. 3L-N.

There exists a qualitative consistence between QPI measurement and DFT simulation at energies explored, as shown in Fig. 3. Detailed analysis of energy-dependent QPI data reveals that these features, such as the crescent moon-shaped and eye-shaped, spiral anticlockwise around the Γ point gradually as energy increases, as shown in Supplementary Information figs. S8 and S9 and the movie M2. The "Fermi arc" state as observed on the (011) surface shows a similar anticlockwise rotation around Γ point as energy increase, as shown in Supplementary Information fig. S5 and movie M1. This is a strong evidence of the Fermi arc's winding around the chiral node as predicted by DFT(*38, 39*).

In the following, we detailedly investigated the winding property of the Fermi arc states. Qualitatively the QPI pattern turns anticlockwise as energy increases. We took a loop-cut along the circle as illustrated in Fig. 4A, and depicted the energy-dependent profiles in Fig. 4B. The QPI dispersive features as marked by yellow and white arrows go up to cross the Fermi energy with chirality. Such chirality can be qualitatively reproduced in the DFT-generated simulation, as shown in Fig. 4C. We emphasize that even though the QPI dispersion doesn't give straightforwardly the band structure, the chirality observed can be surely ascribed to the chirality of the surface Fermi arc states.

Quantitative E-q dispersions of the surface Fermi arc states along high symmetric directions of $\bar{\Gamma}$-$\bar{X}$, $\bar{\Gamma}$-$\bar{X}'$, $\bar{\Gamma}$-$\bar{M}$ and $\bar{\Gamma}$-$\bar{M}'$ as marked in Fig. 4A are also extracted and depicted in Fig. 4D to G. The feature near the center is rather complicated, which should include the mix of multi intra-bulk band scattering at Γ and R and the small vectors from Fermi arc states. Even though to identify the scattering channels is difficult, two Dirac-like band crossings are distinguishable along the specific directions. One of them can be recognized along $\bar{\Gamma}$-$\bar{X}$ and $\bar{\Gamma}$-$\bar{X}'$, located near ~ +50 mV, and another along $\bar{\Gamma}$-$\bar{X}'$ and $\bar{\Gamma}$-M, located at ~ -150 mV, as particularly outlined by the green lines in Fig. 4E and F. They can be assigned to the projected chiral nodes at Γ and R respectively, the energies of which are consistent with the recent

ARPES measurement(*41, 42, 44*). Apart from this, the surface Fermi arc states contribute to the additional dispersive curves as guided by yellow lines in Fig. 4D to G. These yellow lines indicate that the QPI feature appears as a pair, and each of the pair evolves in a similar way, except for an energy shift. On the other hand, at the presence of SOC, one surface state splits into two branches, and thus one would expect to observe the paired QPI features originating from such SOC-split surface states, considering the spin-conserving process. Considering such consistence between the experiment and the SOC split surface states, we believe that the observed paired QPI feature comes from the SOC splitting. The value of SOC splitting varies with momentum and energy. According to Fig. 4, we estimate that the maximum may be up to ~80 mV at ~ +200 mV, in agreement with the DFT calculation(*39*). However, the SOC splitting near Fermi energy is relatively smaller, in the range between ~ 25 mV and ~ 35 mV. Such a SOC-induced Fermi arc splitting is not observed in recent ARPES studies(*41-44*), possibly due to the limit of energy resolution. The surface Fermi arc dispersion also shows strong anisotropy along different directions, as shown in Fig. 4D to G and Supplementary Information fig. S10 and Movie M2, which is originated from the anisotropic contours of the Fermi arcs. The total energy window that the surface Fermi arc spans is estimated as ~ 600 mV (from –200 mV to +400 mV), also consistent with the theoretical expectation.

**CONCLUSIONS**

In summary, through the FFT-STS measurement, we have undoubtedly demonstrated the topological fermi arc states on (001) and (011) surfaces of CoSi single crystals. The observed QPI patterns from these Fermi arcs are consistent with the simulation, which confirms that CoSi hosts the chiral spin-3/2 RSW fermion and double spin-1 Weyl fermions. The chiral surface states exhibit a number of exotic properties, such as the intensity singularity in LDOS, large SOC splitting and wide energy window etc. Further exploration is expected to discover novel physics associated with the surface Fermi arc states.

**Acknowledgements**

**Funding:** This work was supported by the National Natural Science Foundation of China (Grants No. 11774149, 11790311, 11574371, 11574394, 11774423, 11822412 and 11674369), the National Key R&D Program of China (Grants No. 2015CB921000, 2016YFA0300504, 2014CB921103, 2016YFA0300600 and 2018YFA0305700), and the Strategic Priority Research Program of Chinese Academy of Sciences (Grant No. XDB28000000). L.Z. and H.W. are also supported by Science Challenge Project (Grant No. TZ2016004) and K.C.Wong Education Foundation (Grant No. GJTD-2018-01). **Author contributions:** S.L. conceived the project; Q.Y. acquired and analyzed the data with the help of W.Z, C.X, Z.S and Z.J.; L.Z, T.Z., and H.W. performed the DFT calculations; S.T., Y. L., and H.L. grew the single crystals; Z.R., C.T. and Y.S. cut and polished the crystals. Q.Y. and S.L. wrote the manuscript with contributions from all authors. **Competing interests:** The authors declare that they have no competing interests. **Data and materials availability:** All data needed to evaluate the conclusions of the paper are present in the paper and/or the Supplementary Materials. Additional data related to this paper may be requested from the authors.

**Figures and Captions:**

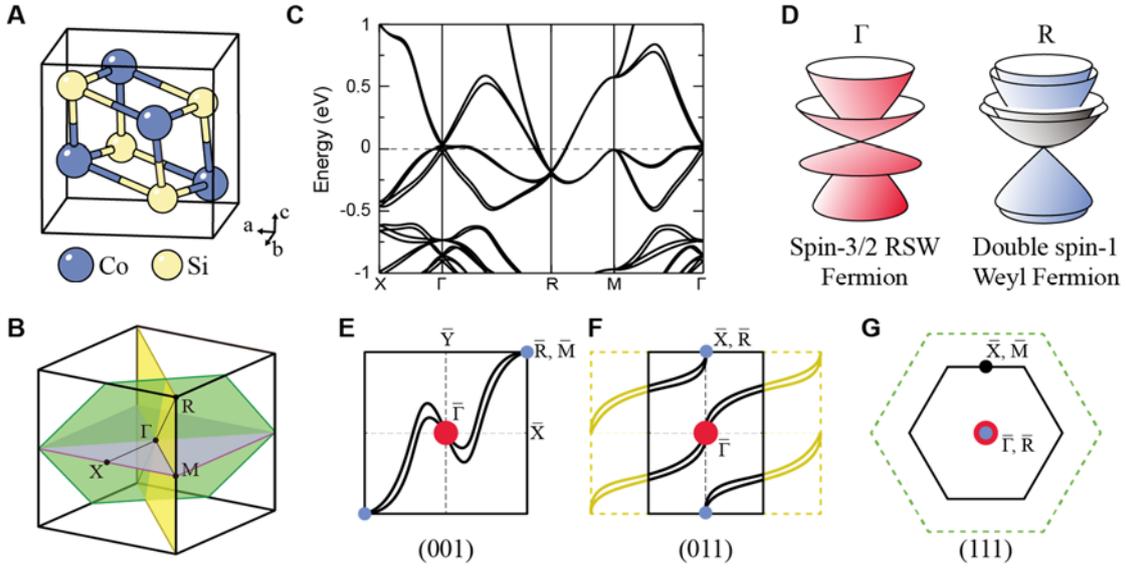

**Fig. 1. Crystal structure and electronic structure of CoSi single crystal.** (**A**) Lattice structure of CoSi (a = b = c = 4.45 Å). The blue and yellow balls represent Co and Si atoms respectively. (**B**) The reciprocal Brillouin zone of CoSi. Γ, X, M and R points are high symmetry positions of the bulk Brillouin zone. The purple, yellow and green planes indicate the projected surface crossing at Γ for the (001), (011) and (111) orientation respectively. (**C**) DFT calculated bulk band structure with spin-orbit coupling along high symmetry directions. (**D**) Schematic illustration of the four-fold degenerated spin-3/2 RSW Fermion node at Γ point and the six-fold degenerated double spin-1 Weyl nodes at R point. (**E** to **G**) Schematic illustrations of the surface Brillouin zones and Fermi arcs for the (001), (011) and (111) surfaces. The colored dashed lines in (**F** and **G**) represent the projected surfaces as shown in (D). The red and blue solid dots indicate the surface projections of Γ and R point respectively. They are located at $\bar{\Gamma}$ and $\bar{M}$ position for (001) surface, and $\bar{\Gamma}$ and $\bar{X}$ position for (011) surface as displayed in (E and F). The black arcs schematically show the surface Fermi arcs connecting the projections of bulk chiral nodes. Γ and R are projected to the same $\bar{\Gamma}$ point on (111) surface.

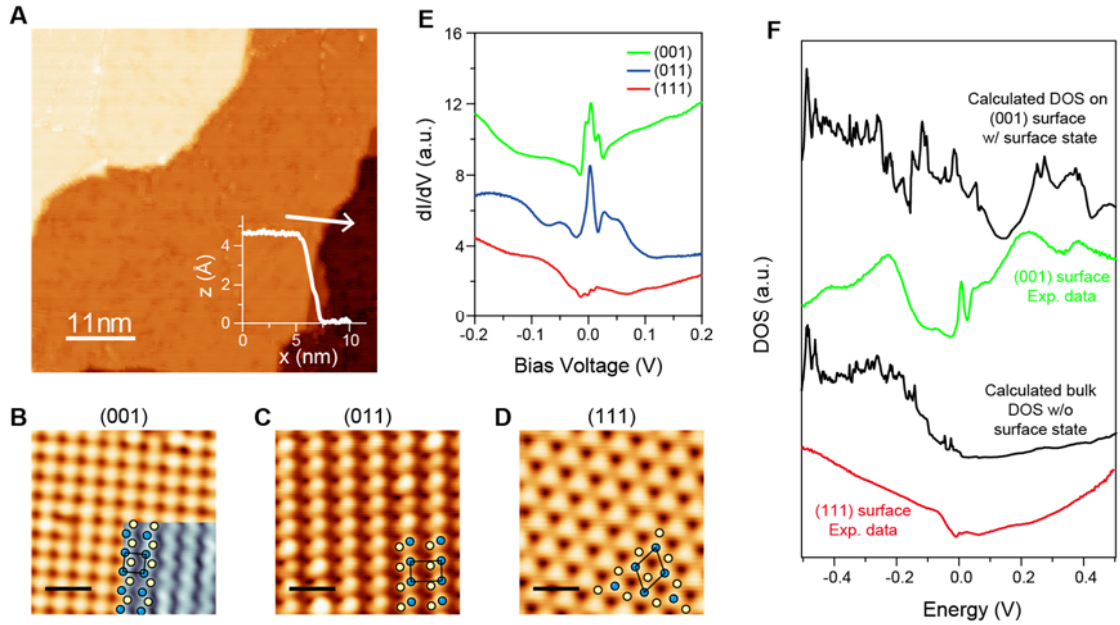

**Fig. 2. STM/STS characterization of CoSi surfaces.** (**A**) Large-scale STM topographic image (55 × 55 nm$^2$, $U$ = 2 V, $I_t$ = 50 pA) of the CoSi (001) surface. Inset: Line-scan profile measured along the white arrowed line, the step height is measured to be ~ 4.5 Å. (**B**) High-resolution STM image (4 × 4 nm$^2$, $U$ = 500 mV, $I_t$ = 100 pA) taken on the (001) surface. Inset: STM image ($U$ = 60 mV, $I_t$ = 1 nA) showing the zig-zag atomic structure. (**C**) High-resolution STM image (4 × 4 nm$^2$, $U$ = -100 mV, $I_t$ = 500 pA) of (011) surface. (**D**) High-resolution STM topography (4 × 4 nm$^2$, $U$ = -500 mV, $I_t$ = 100 pA) taken on the (111) surface. The scale bars for (B to D) are 8 Å, and the corresponding unit cells are also indicated by the square, rectangle, and parallelogram. The blue and yellow dots mark the Co and Si atoms respectively. (**E**) Small-range dI/dV (differential conductance) spectra taken on the three surfaces of CoSi. The green, blue and red curves represent the dI/dV spectrum for (001), (011) and (111) respectively. (**F**) The DFT-calculated (001) projected surface DOS (up black) and experimentally measured dI/dV curve on (001) surface (green), the calculated bulk DOS without surface states (down black), and the measured dI/dV curve on (111) surface (red).

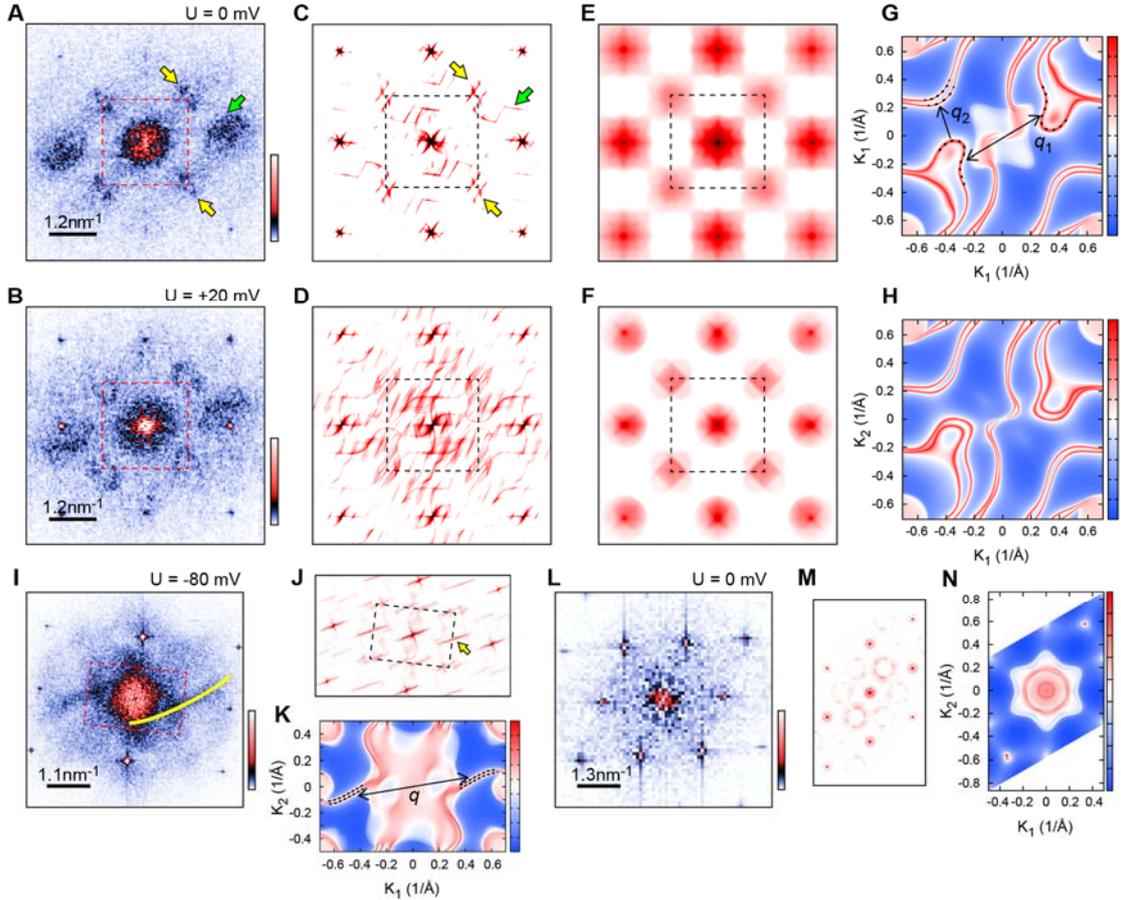

**Fig. 3. Quasiparticle interference (QPI) patterns on various surfaces of CoSi.** (**A** and **B**) FFT images transformed from the dI/dV maps taken on the (001) surface over an area of 25 × 21 nm² at 0 mV and +20 mV. $U$ = +60 mV, $I$ = 200 pA, modulation: 12 mV. The yellow arrows indicate the eye-shaped feature, and the green arrow indicate the crescent moon-like pattern around the Bragg peak. The red dashed square mark the surface first Brillouin zone. (**C** and **D**) QPI simulations including SOC that are generated from both of the surface states and bulk states at $E_F$ (Fermi energy) and $E_F$ +20 mV for (001). There exist two kinds of prominent interference patterns as marked by yellow and green arrows. The black dashed squares mark the first surface BZ, the same as in (A and B). (**E** and **F**) Calculated QPI simulation by removing the surface states part from (C and D) for the (001) surface at $E_F$ and $E_F$ +20 mV, respectively. (**G** and **H**) Calculated constant energy contours at $E_F$ and $E_F$+20 mV of the (001) surface with Fermi arcs traversing the whole Brillouin zone. The black arrows in (G) indicate the corresponding scattering processes of two kinds of prominent QPI patterns in (C). (**I**) FFT image transformed from the dI/dV map taken on (011) surface ($U$ = -80 mV, $I_t$ = 100 pA, modulation: 13 mV) over an area of 40 × 40 nm². The red rectangle marks the first Brillouin zone of (011) surface and the yellow arc highlights the arc-shaped feature. (**J**) QPI simulation including SOC and surface states at -80 mV for (011) surface. The black dashed rectangle marks the first surface Brillouin zone. The yellow arrow indicates the corresponding QPI feature as in (I). (**K**) Calculated constant energy contour for (011) surface at -80 mV. (**L**) FFT image transformed from the dI/dV map taken on (111) surface at 0 mV ($U$ = -60 mV, $I_t$ = 200 pA, modulation: 12 mV) over an area of 10 × 10 nm². (**M**) QPI simulation including SOC and surface states at $E_F$ for (111). (**N**) Calculated constant energy contour for (111) surface at $E_F$.

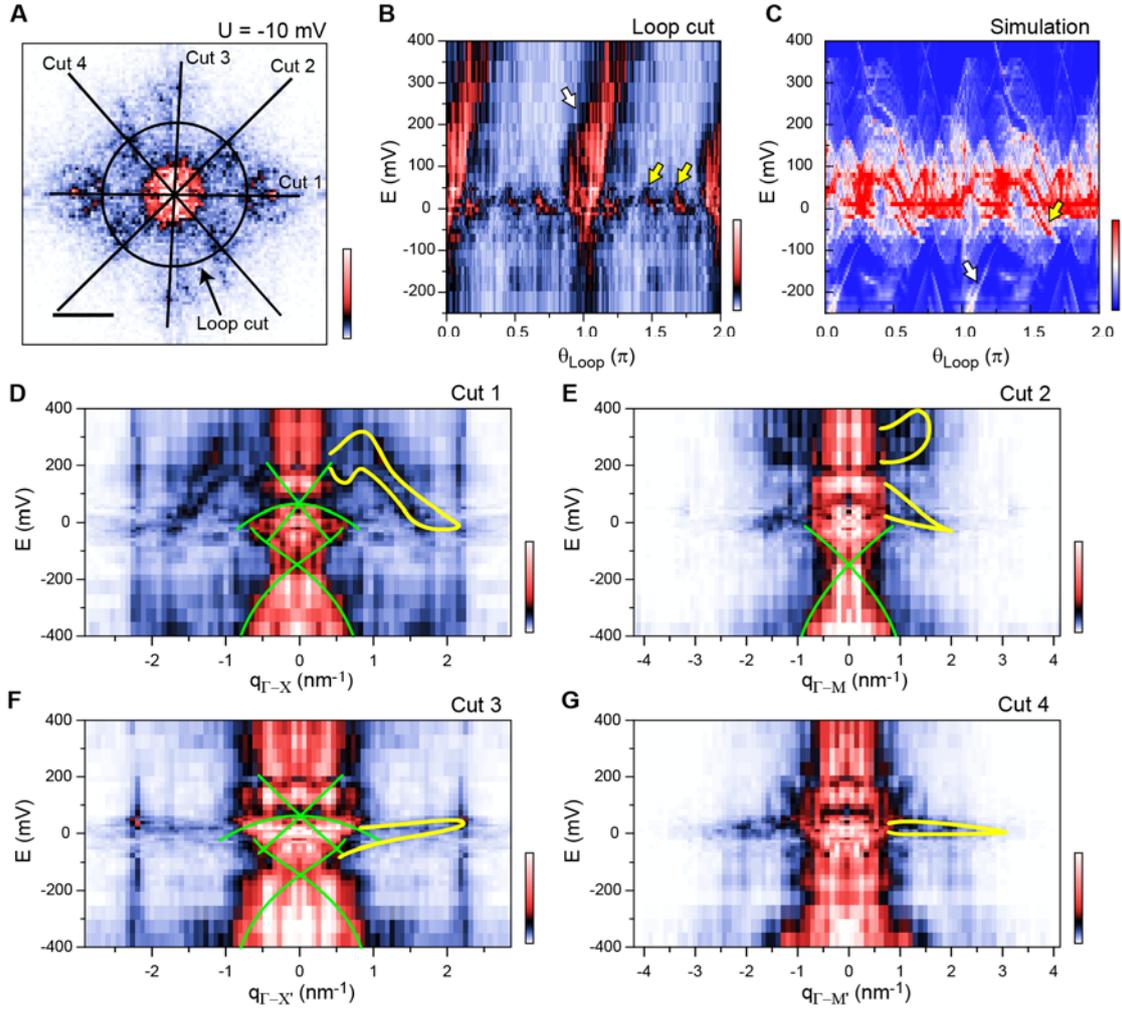

**Fig. 4. Energy dispersion of surface states with chirality.** (**A**) Experimental FFT image ($U$ = -10 mV) transformed from the dI/dV map taken on (001) surface. The scale bar is 1.4 nm$^{-1}$. The black straight lines cuts (1-4) indicate the four high symmetry directions $\bar{\Gamma}$-$\bar{X}$, $\bar{\Gamma}$-$\bar{X}'$, $\bar{\Gamma}$-$\bar{M}$ and $\bar{\Gamma}$-$\bar{M}'$. (**B**) E-q dispersion extracted along the loop-cut as marked by the black circle in (A). The magnitude of the scattering wave vector ($|\vec{q}|$) along the loop is $3\pi/2a$. The θ$_{Loop}$ is rotated anticlockwise and relative to the cut-1. The yellow arrows mark the right chirality of the QPI features along cut-2, cut-3 and cut-4, and the white arrow marks the left chirality of the QPI features along cut-1. (**C**) Calculated E-q dispersion along the loop ($|\vec{q}| = 3\pi/2a$) based on a series of surface states-based QPI simulations. The yellow and white arrows mark the right and left chiral features respectively. (**D** to **G**) E-q dispersions along $\bar{\Gamma}$-$\bar{X}$, $\bar{\Gamma}$-$\bar{X}'$, $\bar{\Gamma}$-$\bar{M}$ and $\bar{\Gamma}$-$\bar{M}'$ as marked in (A) extracted from the energy dependent QPI maps. The green colored guiding lines indicate two Dirac-like band crossings. These two Dirac nodes are located at ~ +50 mV and -150 mV. The additional extensive dispersions of the surface Fermi arcs are marked by yellow colored lines.

## MATERIALS AND METHODS

**Sample preparation.** The CoSi single crystals were grown by chemical vapor transport method. Co and Si powders in 1:1 molar ratio were put into a silica tube with the length of 200 mm and the inner diameter of 14 mm. Then, 200 mg $I_2$ was added into the tube as a transport reagent. The single CoSi crystals with an average size of ~ 2 mm were obtained. Due to the cubic structure and strong covalent bonding, it's difficult to obtain a perfect surface by cleaving single CoSi crystals. So we cut and polished the single crystals before loading into ultrahigh vacuum, and then *in situ* repeatedly $Ar^+$ ion sputtered and annealed the samples until clear RHEED patterns appeared.

**STM and STS characterization.** The STM characterization was performed in a commercial STM system (USM-1500, UNISOKU) at 4.2K. The base pressure is less than $1\times10^{-10}$ Torr). The STM topographic images were collected under constant current mode with a mechanically polished Pt-Ir tip. The dI/dV spectra were acquired with a lock-in amplifier technique.

**DFT calculations.** The calculation based on the density functional theory (DFT) to simulate the electronic structure of CoSi is performed by using the VASP package (*51*), with the generalized gradient approximation (GGA) in Perdew-Burke-Ernzerhof (PBE) form (*52*) as the exchange-correlation functional. The cutoff of energy is set to 450 eV and the Brillouin zone (BZ) integration is performed on $10 \times 10 \times 10$ mesh. Spin-orbit coupling (SOC) is taken into account for all calculations. We build a 10-unit-cell-thick slab to obtain the projected surface local density of states (LDOS), with a 16Å thick vacuum layer along the (001) direction to eliminate the interaction between slabs. In order to investigate the surface states and quasiparticle interference (QPI), a tight-binding model is constructed by the *d* orbits of Co and *p* orbits of Si based on the maximally localized Wannier functions (MLWF) (*53*). The surface states and QPI are calculated by the WannierTools package (*54*).